\documentclass[numberedappendix]{emulateapj}
\usepackage{graphicx}
\usepackage[toc,page]{appendix}
\usepackage{lipsum}

\usepackage{color}
\usepackage{amsmath}

\newcommand\etal{~et~al.}

\begin{document}

\title{New Constraints on Quantum Gravity from X-ray and Gamma-Ray Observations}

\author{E. S. Perlman\altaffilmark{1}, S. A. Rappaport\altaffilmark{2}, W. A. Christiansen\altaffilmark{3}, Y. J. Ng\altaffilmark{3}, J. DeVore\altaffilmark{4}, D. Pooley\altaffilmark{5,6}}
	
\altaffiltext{1}{Department of Physics \& Space Sciences, Florida Institute of Technology, 150 W. University Blvd., Melbourne, FL  32901, USA; eperlman@fit.edu}

\altaffiltext{2}{Department of Physics and Kavli Institute for Astrophysics \& Space Research, Massachusetts Institute of Technology, Cambridge, MA 02139; sar@mit.edu}

\altaffiltext{3}{Department of Physics and Astronomy, University of North Carolina, Chapel Hill, NC  27599, USA; email:  wayne@physics.unc.edu, yjng@physics.unc.edu}

\altaffiltext{4}{Visidyne, Inc., jgdevore@cox.net}

\altaffiltext{5}{Eureka Scientific, 2452 Delmer Street, Suite 100, Oakland, CA  94602-3017; davepooley@me.com}

\altaffiltext{6}{Sam Houston State University; davepooley@me.com}

\begin{abstract}

One aspect of the quantum nature of spacetime is its ``foaminess" at very 
small scales.   Many models for spacetime foam are defined by the accumulation power $\alpha$, 
which parameterizes the rate at which Planck-scale spatial uncertainties (and the 
phase shifts they produce) may accumulate over 
large path-lengths.  Here $\alpha$ is defined by the
expression for the path-length fluctuations, $\delta \ell$, of a source at distance $\ell$, wherein $\delta \ell \simeq 
\ell^{1 - \alpha} \ell_P^{\alpha}$, with $\ell_P$ being the Planck length.  
We reassess previous proposals to use astronomical observations of
distant quasars and AGN to test models of spacetime foam.  We show 
explicitly how wavefront distortions on small scales cause the image 
intensity to decay to the point where distant objects become undetectable when 
the path-length fluctuations become comparable to the wavelength of the radiation.  
We use X-ray observations from {\em Chandra} to set the constraint $\alpha  \gtrsim 0.58$, which rules out 
the random walk model (with $\alpha = 1/2$).  Much firmer constraints can
be set utilizing detections of quasars at GeV energies with {\em Fermi}, and at TeV energies with ground-based Cherenkov
telescopes: $\alpha \gtrsim 0.67$ and $\alpha \gtrsim 0.72$, respectively.  These 
limits on $\alpha$ seem to rule out 
$\alpha = 2/3$, the model of some physical interest.
\end{abstract}


	\maketitle
\eject
	

\section{Introduction}
\label{sec:Intro}

Even at the minute scales of distance and duration examined with
increasingly discriminating instruments, spacetime still appears
to be smooth and structureless. However, a variety of models of 
quantum gravity posit that spacetime is, on Planck scales, subject to
quantum fluctuations. As such, the effect of quantum gravity 
on light propagation (if detected) can possibly reveal a coupling to vacuum 
states postulated by Inflation and String Theories.  
In particular, models (e.g., Ng 2003) consistent with the ``Holographic Principle" 
('tHooft 1993; Susskind 1995; Aharony et al. 2000) predict that space-time foam 
may be detectable via intensity-degraded or blurred images of distant objects.  
While these models are not a direct test of the Holographic Principle itself, the 
success or failure of such models may provide important clues  to connect black 
hole physics with quantum gravity and information theory (Hawking 1975). 

The fundamental idea is that,  if probed at a small enough scale, spacetime 
will appear complicated -- something akin in complexity to a turbulent
froth that Wheeler (1963) has dubbed ``quantum foam,'' also known as 
``spacetime foam.''  In models of quantum gravity, the ÒfoaminessÓ of spacetime 
is a consequence of the Energy Uncertainty Principle connecting the Planck mass 
and Planck time.  Thus, the 
detection of spacetime foam is important for constraining models of 
quantum gravity.   If a foamy structure is found, it would require
that space-time itself has a probabilistic, rather than deterministic nature.  As a 
result, the phases of photons emitted by a distant source would acquire a random 
component which increases with distance.  Furthermore, the 
recent discovery of polarization in the cosmic microwave background 
by BICEP2 (Ade et al.~2014), if confirmed, also provides 
evidence for imprints of quantum gravitational effects from the inflationary
era appearing on the microwave background.  Although these effects 
originate from an epoch vastly different than the present time, they may be
associated with a theoretically expected chaotic (e.g., foamy) inflation of
space-time (for a recent review, see Linde 2014 and references therein).  
Therefore, searching for evidence of quantum
foam in the present era, which is actually slowly inflating because of dark
energy, may also be helpful in providing observational support for theories 
of quantum gravity's role in inflation. 

A number of prior studies have explored the possible image degradation of distant 
astronomical objects due to the effects of spacetime foam (Lieu \& Hillman 2003; 
Ng et al.~2003; Ragazzoni et al.~2003; Christiansen et al.~2006; Christiansen
et al.~2011; and Perlman et al.~2011).  In particular, most of these focus on possible
image blurring of distant astronomical objects. We demonstrate that this previous approach 
was incomplete, and take a different approach, examining the possibility that spacetime foam 
might actually prevent the appearance of images altogether at sufficiently short wavelengths.  
We concentrate particularly on observations with the {\em Chandra X-ray 
Observatory} in the keV range (a possibility we considered unfeasible in Christiansen et 
al.~2011 and Perlman et al.~2011, but now reconsider), 
the {\em Fermi Observatory} in the 
GeV range, and ground-based Cherenkov telescopes in the TeV range.  
Short-wavelength observations are particularly useful in constraining quantum 
gravity models since, in most models of quantum gravity, the path-length fluctuations 
and the corresponding phase fluctuations imparted to the wavefront of the radiation 
emitted by a distant source are given by:
\begin{equation}
\label{eqn:phase}
\delta \phi \simeq 2 \pi \ell^{1-\alpha} \ell_P^{\alpha} / \lambda
\end{equation} 
(Christiansen et al.~2011) where $\lambda$ is the 
wavelength one is observing, the parameter $\alpha \lesssim 1$ specifies different 
space-time foam models, and $\ell$ is the line-of-sight co-moving distance to the source.  
(The prefactor in Eqn.~(\ref{eqn:phase}) may not be exactly $2\pi$, but as we show 
shortly the exact factor is unimportant for the conclusions we draw.)

This paper is organized as follows.  In \S \ref{sec:phase_fluc} we discuss the phase fluctuations 
that might be 
imparted to a wavefront by the spacetime foam, as well as previous attempts to detect them. 
We use several heuristic arguments to derive
the relation for these fluctuations in the context of a `holographic model', as well as other 
models. In \S \ref{sec:phase_effects}
we describe the effects that these phase fluctuations would have on images
of distant astronomical objects, including degrading them to the point where they become
undetectable.  
In \S \ref{sec:high_energy} we utilize well-formed (i.e., $\lesssim 1''$) 
{\it Chandra} X-ray images to constrain the spacetime foam parameter $\alpha$, and 
then move on to set yet tighter constraints on $\alpha$ 
using the lower resolution (i.e., $\sim$$1^\circ$) {\em Fermi}, and even ground-based Cherenkov 
telescope images at TeV energies.  Here the constraints on $\alpha$ 
appear to rule out the value of $\alpha = 2/3$ predicted by the holographic model.  
(For a possible connection to the holographic principle, see \S \ref{sec:holog} where an 
important caveat is also pointed out.)  We summarize our results in \S \ref{sec:summary}.  

\section{The Basic Phase Fluctuation Model}
\label{sec:phase_fluc}

All of the effects discussed in this work depend explicitly on the accumulation power $\alpha$, 
which parameterizes the rate at which minuscule spatial uncertainties, generated at the Planck level 
($\sim$$10^{-33}$ cm), may accumulate over large distances as photons travel through spacetime foam.
Since there is not yet a universally accepted theory of quantum gravity, there is more than one model 
for spacetime foam, so $\alpha$ can, in principle, be treated as a free parameter to 
be determined from observations.  In this picture, the path length fluctuations $\delta \ell$ 
in propagating light beams accumulate according to $\delta \ell \simeq \ell^{1-\alpha} \ell_P^{\alpha}$, 
where  $\ell$, the distance to the source, and $\ell_P$, the Planck length ($\ell_P = \sqrt{\hbar G/ c^3}$),
are the two intrinsic length scales in the problem.  
We note, in passing, that  $\alpha$ bears an inverse relationship with distance, $\ell$, in the sense that 
small values, i.e., $\alpha \rightarrow 0$, correspond to rapidly accumulating fluctuations;  whereas, large 
values of $\alpha \rightarrow 1$ correspond to slow (or even non-existent)  accumulation.  

In spite of the lack of a well-defined model for spacetime foam, some theoretical models for light 
propagation have been developed that specify  $\alpha$ and thereby allow insight into the 
structure of spacetime foam on the cosmic scale.  The most prominent models discussed in 
the literature are:

\begin{enumerate}

\item The random walk model (Amelino-Camelia 1999; Diosi \& Lukacs 1989).  In this model, the 
 effects grow like a random walk, corresponding to  $\alpha=1/2$.

\item The holographic model (Karolyhazy 1966; Ng \& van Dam 1994, 1995),  
so-called because it is consistent (Ng 2003) with the holographic principle (`tHooft 
1993; Susskind 1995). (We explain the meaning of ``consistent" below.)
In this model, the information content in any three-dimensional 
region of space can be encoded on a two-dimensional surface surrounding the region
of interest, like a hologram. (This is the restricted form of the holographic principle that
we are referring to.)  The holographic model corresponds to a value of
$\alpha = 2/3$ (Christiansen et al.~2011).  \ 

\item The original Wheeler conjecture, which in this context means that the distance 
fluctuations are anti-correlated with successive fluctuations (Misner et al.~1973), in which 
case there are no cumulative effects, so that the distance fluctuation remains simply the 
Planck length.	This corresponds to $\alpha=1$ and spacetime foam is virtually
undetectable by astronomical means.

\end{enumerate}

While all three of the above models  are  tested by the techniques discussed below,  
we devote most of our attention in this paper to the holographic model (\# 2 above) because 
it is most directly connected to theories of quantum gravity via the holographic principle.

\subsection{A Short Review of the Holographic Model}
\label{sec:holog}

To understand how large the quantum fluctuations of spacetime are (as reflected by the
fluctuations of the distances along a null geodetic path) in the holographic model 
(Ng \& van Dam 1994; Karolyhazy 1966), let us consider mapping
out the geometry of spacetime for a spherical volume of radius $\ell$ over the
amount of time $2\ell/c$ it takes light to cross the volume (Lloyd \& Ng 2004).  
One way to do this is to fill the space with clocks, exchanging
signals with the other clocks and measuring the signals' times of arrival. 
The total number of operations, including the ticks of the clocks and
the measurements of signals, is bounded by the Margolus-Levitin
theorem (Margolus \& Levitin 1998) which stipulates that the rate of operations 
cannot exceed the amount of energy $E$ that is available for the operation 
divided by $\pi \hbar/2$.  A total mass $M$ of clocks then
yields, via the Margolus-Levitin theorem, the bound on the total number of
operations given by $(2 M c^2 / \pi \hbar) \times 2 \ell/c$. But to prevent
black hole formation, $M$ must be less than $\ell c^2 /2 G$. Together, these
two limits imply that the total number of operations that can occur in a
spatial volume of radius $\ell$ for a time period $2 \ell/c$ is no greater than
$2 (\ell/l_P)^2 / \pi$. To maximize spatial resolution, each clock must tick
only once during the entire time period. If we regard the operations
as partitioning the spacetime volume into ``cells", then on the average each cell
occupies a spatial volume no less than $\sim \ell^3 / (\ell^2 / \ell_P^2)
= \ell \ell_P^2$, yielding an average separation between neighboring
cells no less than $ \sim \ell^{1/3} \ell_P^{2/3}$ (Ng 2008).
This spatial separation can be interpreted as the average minimum uncertainty in the
measurement of a distance $\ell$, that is, $\delta \ell \gtrsim \ell^{1/3}
\ell_P^{2/3}$.

An alternative way to derive $\delta \ell$ is to consider the Wigner-Salecker gedanken experiment.
In the Wigner-Salecker experiment (Salecker \& Wigner 1958; Ng \& van Dam 1994)
a light signal is sent from a clock to a mirror (at a distance $\ell$ away) and back to the clock 
in a timing experiment to measure $\ell$. The clock's and the mirror's positions, according to 
Heisenberg's uncertainty principle, will have a positional uncertainty of $\delta \ell$; the uncertainty
in the clock's position alone, implies $(\delta \ell)^2 \gtrsim \hbar \ell / mc$, where $m$ is the 
mass of the clock.  Now consider
the clock to be light-clock consisting of a spherical cavity of diameter $d$,
surrounded by a mirror wall between which bounces a beam of light. For the
uncertainty in distance not to exceed $\delta \ell$, the clock must tick off time
fast enough so that $d/c \lesssim \delta \ell /c$. But $d$ must be larger than
twice the  Schwarzschild radius $2Gm/c^2$. These two requirements imply 
$\delta \ell \gtrsim 4Gm/c^2$ (Ng \& van Dam 1994; Karolyhazy 1966) 
to measure the fluctuation of a distance $\ell$. This latter expression for $\delta \ell$ can 
be multiplied with the above constraint equation based on the requirement from quantum
mechanics to yield $(\delta l)^3 \gtrsim 4 \ell \ell_P^2$ (independent of the mass $m$ of the clock). 
We conclude that the fluctuation of a distance $\ell$ scales as $\delta \ell \gtrsim \ell^{1/3} \ell_P^{2/3}$.

The following heuristic argument may help to explain why we interpret the result $\delta \ell \gtrsim \ell^{1/3} \ell_P^{2/3}$
as being consistent with the holographic principle.  First recall that the
holographic principle ('tHooft 1995, Wheeler 1982; Bekenstein 1973; Hawking 1975; Susskind 1995)
states that the maximum number of degrees of
freedom that can be put into a region of space is given by the area of the
region in Planck units.  Consider a
region of space measuring $\ell \times \ell \times \ell$, and imagine partitioning it
into cubes as small as physical laws allow.  With each small cube we
associate one degree of freedom.   If the smallest uncertainty in
measuring a distance $\ell$ is $\delta \ell$, in other words, if the fluctuation
in distance $\ell$ is $\delta \ell$, then the smallest such cubes have volume
$(\delta \ell)^3$.  (Otherwise, one could divide $\ell$ into units each measuring
less than $\delta \ell$, and by counting the number of such units in $\ell$, one
would be able to measure $\ell$ to within an uncertainty smaller than
$\delta \ell$.)  Thus the maximum number of degrees of freedom,
given by the number of small cubes we can put into the region of
space, is $(\ell/ \delta \ell)^3$.  It follows from the holographic principle
that $(\ell / \delta \ell)^3 \lesssim (\ell / \ell_P)^2$, which yields precisely the
result $\delta \ell \gtrsim \ell^{1/3} \ell_P^{2/3}$.  It is in this sense that
our so-called holographic spacetime foam model is consistent with
the holographic principle -- no less and no more.  In spite of this apparent consistency,
we call the readers' attention to this important caveat: ruling out the $\alpha = 2/3$ holographic
model does {\it not} necessarily imply the demise of the holographic principle, for
the correct spacetime foam model associated with the holographic principle may take on a
different and more subtle form than that which can be
given by $\delta \ell \sim \ell^{1/3} l_P^{2/3}$.

\subsection{A Short History of Attempts to Detect Spacetime Foam}

To assist the readers in placing our discussion in proper context, let us provide
a brief (necessarily incomplete) history of the various proposals to
detect spacetime foam models. Among the first proposals was the use of 
gravitational wave interferometers (such as LIGO, VIRGO and LISA) to measure 
the foaminess of spacetime which is expected to provide a (new) source of noise 
in the interferometers (Amelino-Camelia 1999; Ng \& van Dam 2000). Implicit in this 
proposal is the assumption that space-time in between the mirrors in the interferometer 
fluctuates coherently for all the photons in the beam.  But the large beam size in LIGO 
and similar interferometers (compared to the Planck scale) makes such coherence 
unlikely.

Another proposal was to 
attribute energy threshold anomalies encountered in the ultra-high energy 
cosmic ray events (at $\sim$$10^{19}$ eV; see, e.g., Lawrence et al.~1991) 
and the 20 TeV-$\gamma$ events (e.g., from Mkn 501; see, e.g., Aharonian et al.~1999 
and Harwit et al.~1999) to energy-momentum uncertainties due to quantum gravity effects 
(Amelino-Camelia \& Piran 2001; Ng et al. 2001).  

Then the possibility of using spacetime foam-induced
phase incoherence of light from distant galaxies and gamma-ray bursts to
probe Planck-scale physics was put forth (Lieu \& Hillman 2003;
Ragazzoni et al.~2003; Ng et al.~2003)\footnote{It was pointed out by Ng et al.~(2003) 
that both Lieu \& Hillman (2003) and Ragazzoni et al.~(2003) did not utilize the correct accumulation factor.}.
It was then pointed out that 
modern telescopes might be on the verge of testing theories of spacetime
foam (Christiansen et al.~2006; Steinbring 2007). The essence of these proposals was a 
null test; i.e., since many theories of spacetime foam predict ``blurring'' of images of distant 
point sources, the absence of deviations from a given telescope's ideal PSF would provide 
evidence for rejecting such theories.   As mentioned above, since the effects of spacetime 
foam on light propagation are so tiny, accumulation over large distances is a necessary 
prerequisite for the viability of any theory.  In this regard, sources (e.g., quasars, blazars, etc.)~at cosmological 
distances would be the preferred targets and the importance of using the appropriate 
distance measure (viz., the line-of-sight comoving distance) of the distant sources for calculating the
expected angular broadening was emphasized (Christiansen et al.~2011)\footnote{The difference
in using the luminosity distance (Steinbring 2007; Tamburini et al.~2011) versus
the line-of-sight comoving distance is significant (Perlman et al. 2011).}.

All of the previous workers who envisioned using images of cosmologically distant objects 
to detect evidence of spacetime foam also adopted the additional hypothesis that the rms 
phase fluctuations, $\delta \phi$, might also be directly interpretable, to within the same order 
of magnitude, as the angular diameter of a spacetime foam induced ``seeing disk" for a distant 
source, $\delta \psi$ -- which we now believe is not justified (see \S\ref{sec:ripples} and \S\ref{sec:phase}).  

Last but not least, time lags from distant pulsed sources such as gamma ray bursts
were posited as a
possible test of quantum gravity (Amelino-Camelia et al. 1998).  This spread in
arrival times from distant sources was found to depend on the energies of the photons
in some formulations of quantum gravity. Indeed super-GeV photons for the {\it Fermi}-detected
Gamma-ray bursts (Abdo et al. 2009) could be used to yield tight bounds
on light dispersion (Nemiroff et al. 2012).  However, when
applied to spacetime foam models parametrized by
$\delta \ell \simeq \ell^{1 - \alpha} \ell_P^{\alpha}$, such time lags were
shown to be energy-independent
and to yield rather small effects  (Ng 2008; and explicitly demonstrated in
Christiansen et al. 2011 for the {\it Fermi}-detected GRBs) due to the equal
probability of positive and negative fluctuations in the speed of light
inherent in such models.


\section{Effects of the Putative Spacetime Foam on Astronomical Images}
\label{sec:phase_effects}

With the description of \S 2 as our backdrop, we now 
take a fresh look  at the effects that the hypothesized spacetime foam may
have on images of distant point-like astronomical objects.  As discussed in the 
Introduction, there are good reasons to believe that spacetime foam would produce
small phase shifts in the wavefronts of light arriving at telescopes.
We first examine quantitatively the effects on astronomical images due to the
expected phase shifts as a function of the parameter, $\alpha$ (\S~\ref{sec:ripples}).  
We then carry out a variety of simulations related to the subject that are described in \S~\ref{sec:phase}.  
As a result of this, we will see that all previous work on using observations 
of distant objects to detect spacetime foam needs to be reformulated.  We accomplish  this
in \S\ref{sec:constraints}.
 
\subsection{Effects of Phase Ripples on the Wavefront of a Distant Astronomical Object}
\label{sec:ripples}

According to  Eqn.~(\ref{eqn:phase}), the fluctuations in the 
phase shifts over the entrance aperture of a telescope or interferometer are described by
\begin{equation}
\Delta \phi(x,y) \simeq 2 \pi \frac{\ell^{1-\alpha} \ell_P^\alpha}{\lambda}
\end{equation}
where $\{x,y\}$ are  coordinates within the aperture at any time $t$,  
$\ell$  is the line-of-sight comoving distance to the source (see discussion in Christiansen et al.~2011), 
and $\ell_P$ is the Planck length.  Given that the Planck scale is extremely small, we
envision that $\Delta \phi(x,y)$ can be described by a random field with rms scatter
\begin{equation}
\delta \phi_{\rm rms} \simeq 2 \pi \frac{\ell^{1-\alpha} \ell_P^\alpha}{\lambda}
\label{eqn:phirms}
\end{equation}
without specifying exactly on what scale in the $x-y$ plane these phase distortions are 
correlated (if any).  However, for purposes of this work we assume that the fluctuations
are uncorrelated down to very small scale sizes $\delta x \times \delta y$, perhaps even 
down to the Planck scale itself.

To help understand what images of distant, unresolved sources might look
like after propagating to Earth through an effective `phase screen' (due to 
spacetime foam) consider an idealized telescope of aperture $D$ forming the image.
Conceptually, the potential image quality contained in the information carried by the
propagating wave is independent of whether an actual telescope forms the image, 
but rather depends only on the phase fluctuations imparted to the wavefront.  
Nonetheless, conceptually, it is easier to think of a conventional set of optics 
forming the image.

In that case, the image formed is just the absolute square of the Fourier transform
of the aperture function, specifically the Fourier transform of $e^{i \Delta \phi(x,y)}$
over the coordinates $\{x,y\}$ of the entrance aperture.  This complex phase screen
can be broken up into two parts as:
\begin{equation}
P(x,y) = \cos \Delta \phi(x,y) + i \sin \Delta \phi(x,y)
\label{eqn:phasescreen}
\end{equation}
and $\Delta \phi(x,y)$ can be considered to be, in our picture of spacetime foam, a
random field with a certain rms value of $\Delta \Phi \equiv \delta \phi_{\rm rms}$.
For small $\Delta \Phi$, Eqn.~(\ref{eqn:phasescreen}) can be written as $P(x,y) \simeq 1+i\Delta \phi(x,y)$.
The Fourier transform of this evaluated over the aperture is the Airy disk function, 
$4J_1(\pi \theta D/\lambda)^2/(\pi \theta D/\lambda)^2$,
with a small amount of white noise superposed (note that here $J_1$ is the 
order 1 Bessel function of the first kind, and $\theta$ is the angular offset from the position
of the source).  At the opposite extreme, for very large
values of $\Delta \Phi$, both the real and the imaginary parts of $P$ fluctuate randomly
between $-1$ and $+1$ with no correlations from point to point within the 
aperture.  The Fourier transform of such a white noise field is just Gaussian white noise.
In other words, no image is formed, and the radiation is dispersed in all directions.

\begin{figure}
\centering
{\includegraphics[width=0.99\columnwidth]{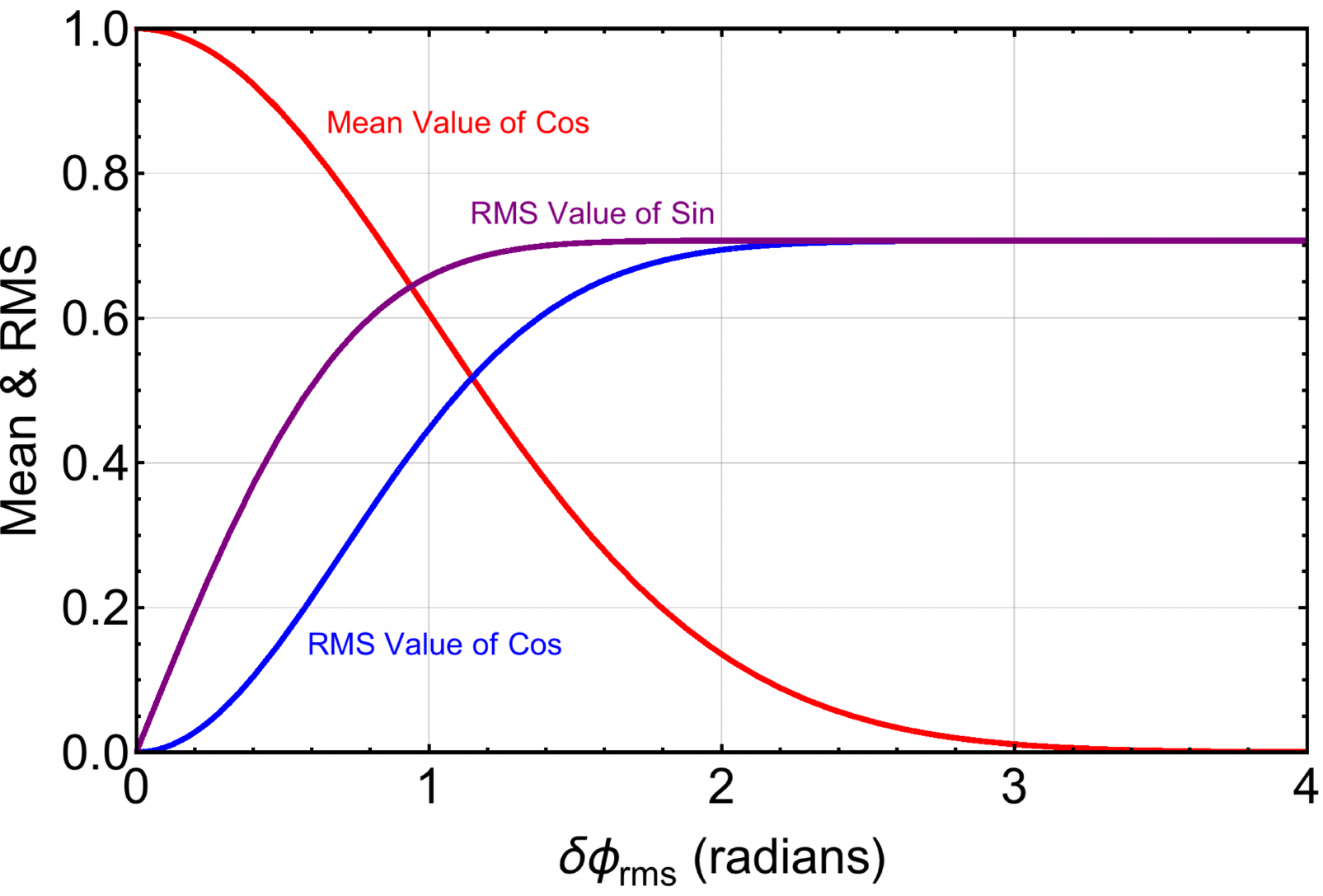}}
\caption{Plots of the expressions given in Eqns.~(5) and the square root of expressions (7) and (8).  The red curve is the expression (5) whose square is the amplitude of the Airy disk function that emerges of the point source in the image. The blue and purple curves are plots of the square root of expressions (7) and (8) which dictate the real and imaginary parts of the underlying (or overlying) white noise due to the phase fluctuations in the wavefront.  The sum of the squares of all three curves equals unity for all values of $\Delta \Phi =\delta \phi_{\rm rms}$.} 
\label{fig:degrade}
\end{figure}

The question of how the image degrades with increasing rms phase fluctuations,
$\Delta \Phi$, can be answered by computing the mean and rms values of $\cos \Delta \phi(x,y)$
and $\sin \Delta \phi(x,y)$ for a given assumed distribution in $\Delta \phi$ with rms value $\Delta \Phi$.
The answers for an assumed Gaussian distribution in $\Delta \phi$ are analytic:
\begin{eqnarray}
\label{eqn:degrade}
\langle  \cos \Delta \phi(x,y) \rangle & =&  e^{-\Delta \Phi^2/2} \\
\langle  \sin \Delta \phi(x,y) \rangle & =&  0 \\
\langle  \cos^2 \Delta \phi \rangle - \left(\langle \cos \Delta \phi \rangle \right)^2 & =&  e^{-\Delta \Phi^2} \left[\cosh \Delta \Phi^2 - 1\right] \\
\langle  \sin^2 \Delta \phi \rangle  & =&  e^{-\Delta \Phi^2} \sinh \Delta \Phi^2  
\end{eqnarray}
Figure \ref{fig:degrade} shows plots of Eqn.~(5), and the square roots of Eqns.~(7) and (8), i.e., the rms values of the $\sin$ and $\cos$ terms, respectively.  The Fourier transform (squared) of the constant term represented by Eqn.~(5) yields the Airy disk function, but with a degraded amplitude given by $e^{-\Delta \Phi^2}$.  By contrast, the Fourier transforms (squared) of the randomly fluctuating parts of the $\sin$ and $\cos$ terms yield a constant white noise background superposed on the degraded amplitude of the Airy disk.  The plots in Fig.~\ref{fig:degrade} show how the Airy disk decays and the white noise increases as a function of the rms amplitude of the phase fluctuations $\Delta \Phi=\delta \phi_{\rm rms}$.

What this demonstrates is that, as the rms amplitude of the phase fluctuations increases, the Airy disk function representing the point source is degraded in amplitude but not in shape, and there is an ever increasing background of white noise superposed.  Since that white noise is essentially spread over all angles in the image plane, the image of the point source simply and effectively decays to the point where it blends in with whatever other instrumental or sky background dominates. Ultimately, when $\Delta \Phi$ approaches $\pi$ radians, the image would simply vanish.  The vanishing of the image results from the complete de-correlation of the wave by destructive interference caused by the large phase fluctuations.

\begin{figure}
\centering
\includegraphics[width=0.99\columnwidth]{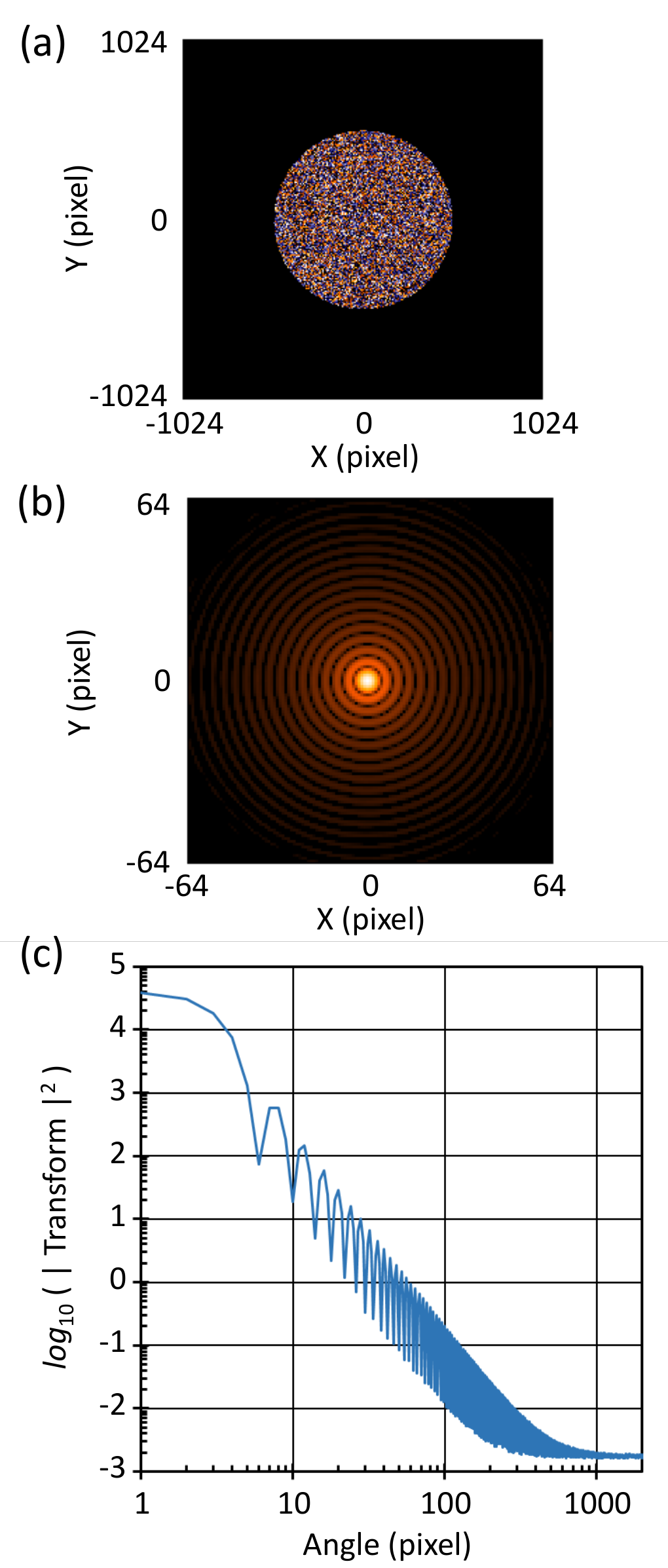}
\caption{Illustrative example of our numerical simulations of a point spread function that 
has been affected by a Gaussian random field of phase shifts over the aperture. The upper 
panel shows the circular aperture with the phase shifts that form a Gaussian field with
an rms amplitude of 0.03 $\lambda$. The circular aperture is embedded in an opaque
screen that is $4096 \times 4096$ pixels, and only the central $2048 \times 2048$ pixels
are shown. The middle panel is the absolute square of the Fourier transform of the 
aperture function displayed using a 1/4-power color palette.  The middle panel shows 
only the central $128 \times 128$ pixels, allowing the Airy rings to be seen easily.   
The lower panel shows a plot of the angularly averaged radial profile
of the absolute square of the FT.}
\label{fig:PSF1}
\end{figure}

\begin{figure*}
\centering
\includegraphics[width=0.8\textwidth]{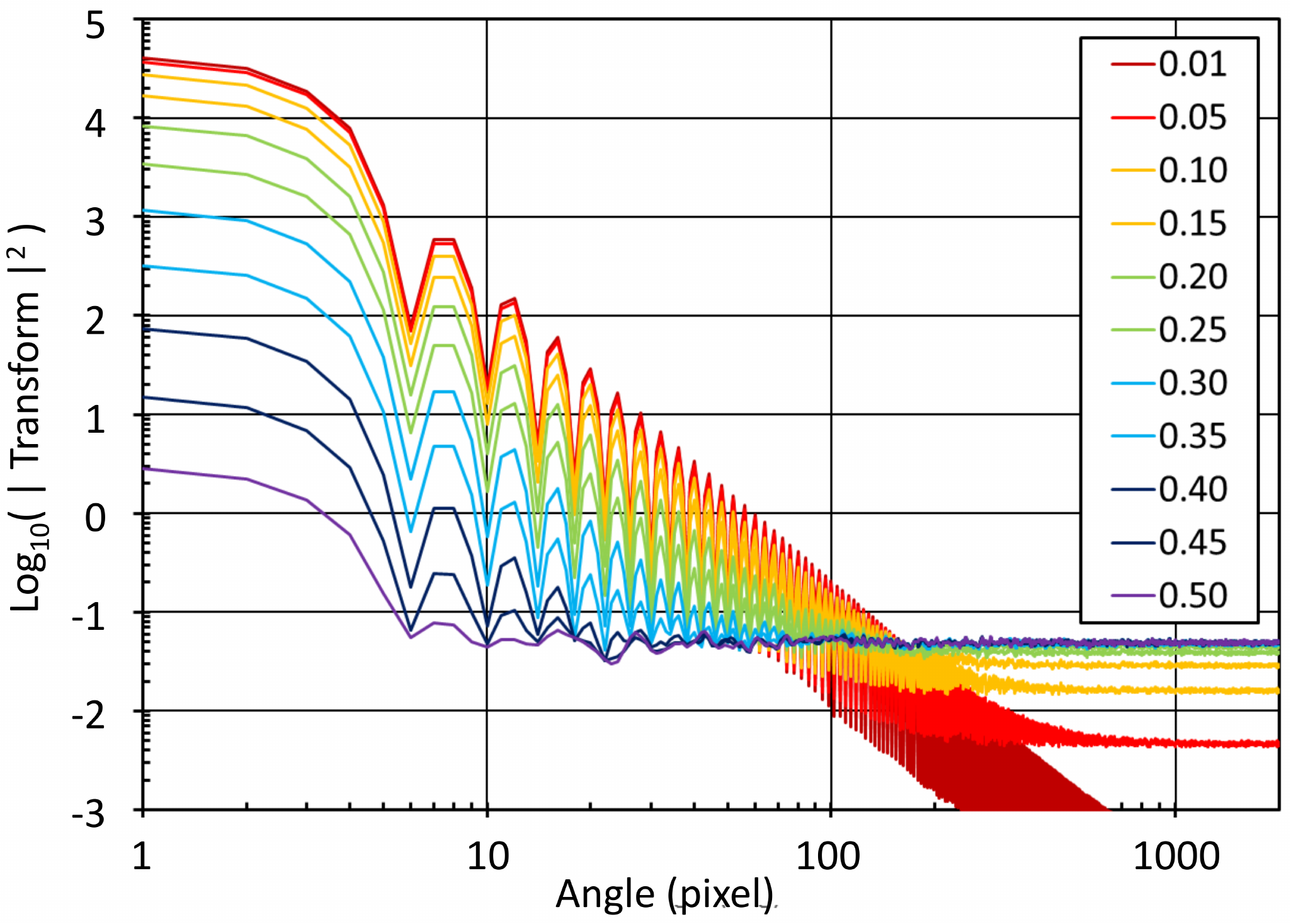}
\caption{Sequence of radial profiles of the numerically computed PSFs for a progression
of rms amplitudes of the phase shifts (assumed to be Gaussian random fields).  The rms
phase shifts range from 0.01 $\lambda$ to 0.5 $\lambda$, as indicated by the color coding.   Note how the {\em shape} of the PSF for small angles is nearly unchanged
until it plateaus into the background.} 
\label{fig:PSF2}
\end{figure*}

\subsection{Degradation of Images Due to `Phase Screens'}
\label{sec:phase}

We have demonstrated in 
\S 3.1 that as long as $\delta \phi_{\rm rms} \lesssim 0.6$ radians
(or $\delta \ell_{\rm rms} \lesssim 0.1 \lambda$; as we show below), 
then the Strehl ratio, which measures the ratio of the peak in the point spread function (`PSF')
compared to the ideal PSF for the same optics, is to a good approximation
\begin{equation}
S \simeq e^{-\delta \phi_{\rm rms}^2} ~~.
\label{eqn:Strehl}
\end{equation}
In addition, if these phase shifts are distributed randomly over the aperture (unlike 
the case of phase shifts associated with well-known aberrations, such as coma,
astigmatism, etc.) then the {\em shape} of the PSF, after the inclusion of the phase shifts
due to the spacetime foam is basically unchanged, except for a progressive decrease 
in $S$ with increasing $\delta \phi_{\rm rms}$ -- as we show next.  

\begin{figure}
\centering
\includegraphics[width=0.99\columnwidth]{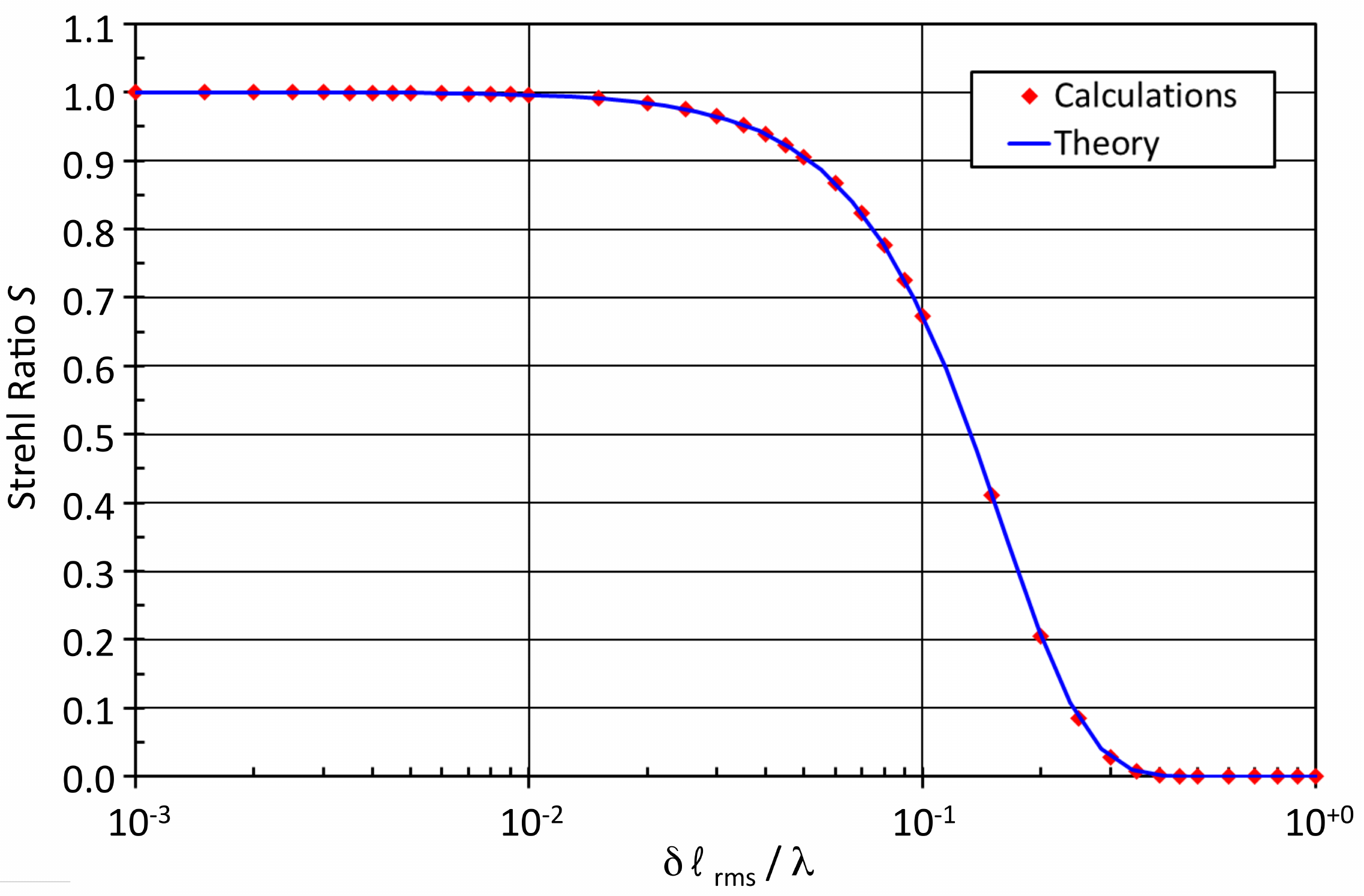}
\caption{Strehl ratio (red points) computed from the numerical simulations of the
PSF as a function of the rms amplitude of the phase shifts (in units of $\lambda$).  The blue
curve is the Gaussian approximation to the Strehl ratio (see Eqn.~3).} 
\label{fig:Strehl}
\end{figure}

We have carried out numerous numerical simulations utilizing various random fields $\Delta \phi(x,y)$,
including Gaussian, linear, and exponential.  We considered a large range of rms values and 
different correlation lengths within the aperture.
Our simulations consisted of a circular aperture that is 1024 pixels in diameter, embedded
in a square array of $4096 \times 4096$ pixels.  The type of calculation we have carried
out is illustrated in Fig.~\ref{fig:PSF1}.  We show the aperture function with an Gaussian 
distribution\footnote{Note, however, that based on the central limit theorem, the results will 
hold for essentially any distribution of phase fluctuations with well-defined rms variations.} 
of random phase fluctuations in the upper panel.  The middle panel shows the 
absolute square of the 2D Fourier transform of this aperture/phase function.  
We then take the results from the middle panel and plot the azimuthally averaged 
radial profile in the lower panel.    

Figure \ref{fig:PSF2} shows a sequence of these PSFs, in the form of radial profiles,
for a range of increasing amplitudes of random phase fluctuations.  As can
be seen, there are three major effects: (i) the peak of the 
PSF is decreased; (ii) beyond a certain radial distance, the PSF reaches a noise 
plateau that can be interpreted as an indication of the partial de-correlation of the wave caused by increasing phase fluctuations; 
and (iii) in between, the shape (including the slope, intensity ratios of Airy rings, etc.) 
of the PSF is {\it unchanged} by the increasing phase fluctuations. The self-similar invariance of 
the PSF shape (aside from the appearance of the noise plateau) contradicts the expectation 
from previous work (e.g., Ng et al. 2003; Christiansen et al.~2006, 2011; Perlman et al. 2011; 
Lieu \& Hillman 2003; Ragazzoni et al.~2003; Steinbring 2007; Tamburini et al.~2011) that phase 
fluctuations could broaden the  apparent shape of a telescope's PSF; thus, allowing for tests of 
spacetime foam models via the Strehl ratio at a level where $\delta \ell/\lambda \approx \lambda/D \ll \pi$. 
In contrast, we now find that for the above criterion, the images are essentially unaffected, while for 
sufficiently large amplitude phase fluctuations (e.g., $\delta \ell/\lambda \gtrsim \pi$) the entire 
central peak ultimately disappears and the image is undetectable, as we showed in \S \ref{sec:ripples}.

\begin{figure*}
\centering
{\includegraphics[width=1.6\columnwidth]{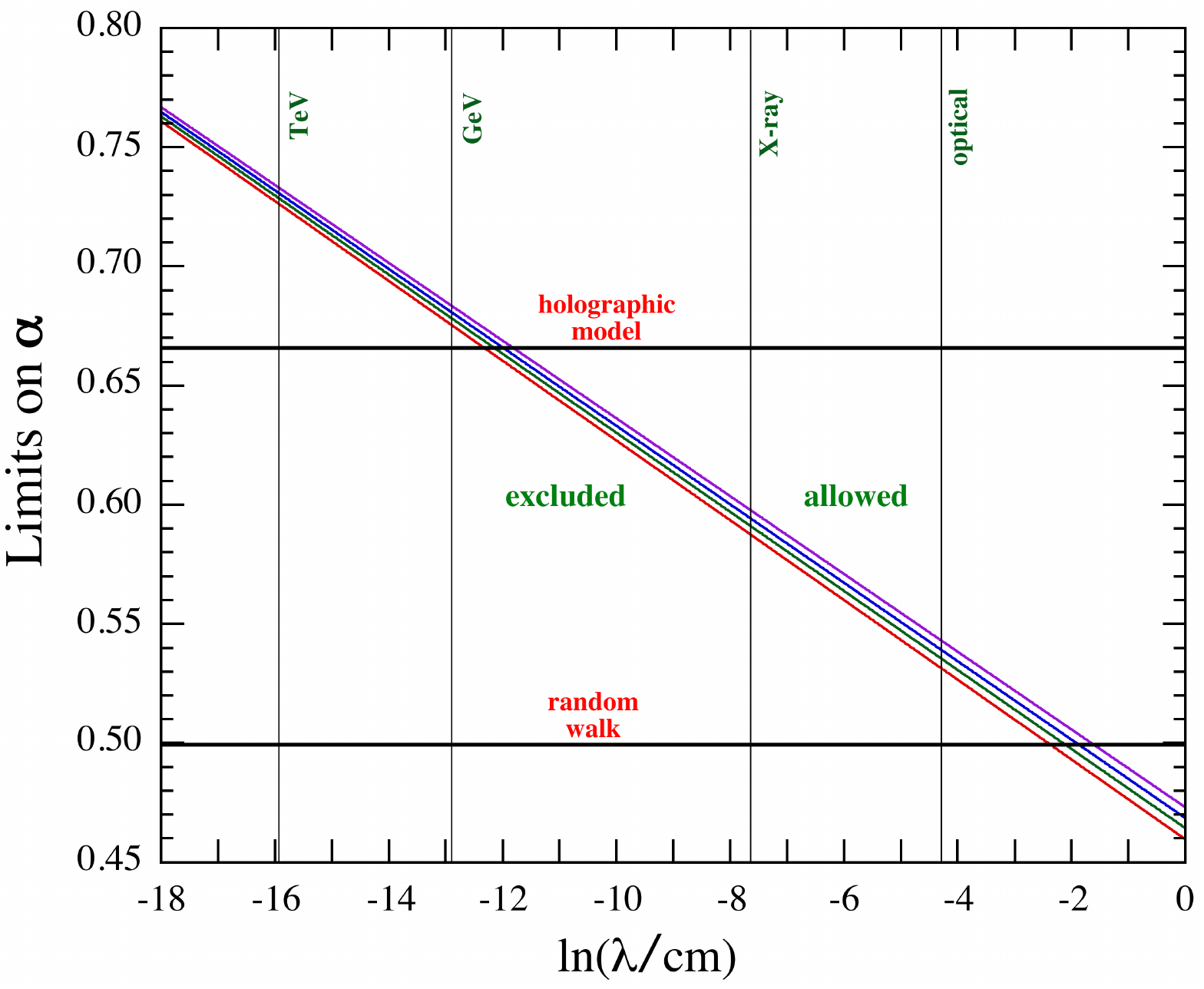}}
\caption{Constraints on the \ parameter $\alpha$, for four different comoving distances
to the object, respectively 300 Mpc ($z\approx 0.07$; red curve), 1 Gpc ($z\approx0.25$; green), 
3 Gpc ($z\approx 1$; blue) and 10 Gpc ($z\approx 12$; purple).  The two horizontal lines refer 
to the holographic and random-walk models, respectively, as labeled.  The vertical dashed lines 
represent the optical (5000 \AA), X-ray (5 keV), GeV and TeV wavebands.  As astronomical images
betray no evidence of cosmic phase fluctuations that might be due to spacetime foam, the region of
parameter space excluded by observations in each band lies below the curves. 
For any given wavelength, $\lambda$, images will not propagate for values of $\alpha$ below the 
various lines corresponding to different comoving distances. } 
\label{fig:alpha}
\end{figure*}

Finally, in Figure \ref{fig:Strehl}, we show a summary plot of the Strehl ratio, as computed from the
numerical simulations, as a function of the rms phase fluctuations (expressed in units
of $\lambda$).  The superposed curve is just a plot of the approximate analytic expression
for the Strehl ratio given by Eqn.~(\ref{eqn:Strehl}).  Not surprisingly, the match is essentially
perfect. The essential point to note here is that the peak of the image ranges from a very 
large fraction of its maximum possible intensity to essentially vanishing as $\delta \ell/\lambda$ 
varies by merely a factor of $\sim$5.

Therefore, since we do not know the intrinsic luminosity of distant quasars we cannot use the
Strehl ratio itself to set constraints on the degree of rms phase fluctuations due to the
intervening spacetime foam.   Indeed, as Figure \ref{fig:PSF2}  shows, the overall PSF shape and
the slope of its decline is nearly unchanged until the phase differences imposed by spacetime foam approach a radian,
at which point the profile just merges into the background noise floor. 
As can be seen in Figs.~\ref{fig:PSF2} and \ref{fig:Strehl}, there is little change in the PSF amplitude until 
$\delta \ell$ gets to within a factor of $\sim$5 of $\lambda$, after which the amplitude plummets.
All that we are able to conclude, is that if $\delta \phi_{\rm rms}$ 
exceeds a certain critical value of $\sim$$\pi$ radians, then the quasar intensity would basically be 
degraded to the point where it would no longer be detected. 

\subsection{Re-Conceptualizing How $\alpha$ Can Be Constrained}
\label{sec:constraints}

The above work allows us to invert Eqn.~(\ref{eqn:phirms}) to set a generic constraint on $\alpha$ for 
distances, $\ell$, to  remote objects as a function of the wavelength,
$\lambda$, used in the observations.  What we find is that 
\begin{equation}
\label{eqn:constraint}
\alpha > \frac{\ln(\pi \ell/\lambda)}{\ln(\ell/\ell_P)},~~
\end{equation} 
{\noindent where we have required a phase dispersion $\delta \phi_{rms} = 2$ radians, corresponding to the
location where the Strehl ratio in Fig. 4 has fallen to $\sim$2\% of its full value. 
We show in Fig.~\ref{fig:alpha} a plot 
of the limit that can be set on the parameter $\alpha$ as a function of measurement 
wavelength, for four different values of comoving distance.  The result is an
essentially universal constraint that can be
set simply by the detection of distant quasars as a function of the observing wavelength.
We shall discuss the effect of this more rigorous understanding of the constraints one can
set on $\alpha$ using observations in X-rays  and $\gamma$-rays in  \S
\ref{sec:Xray} and \S \ref{sec:gamma} respectively. However, in the optical, contrary to 
previous works (including our own), the constraint on $\alpha$ 
is now found to be only $\alpha> 0.53$, i.e., ruling out the random walk 
model, but not coming close to the parameter space required for the holographic model.

A second, completely equivalent way to think of this constraint, is to point out that the 
$\alpha$-models predict that at any wavelength,  $\lambda$, spacetime foam sets a 
maximum distance, beyond which it would simply be impossible to detect a cosmologically 
distant source.  To demonstrate this, we show in Fig \ref{fig:distmax} a plot of the relative 
flux density $\nu F_\nu$ with which a source would be detected, as a function of wavelength.  This plot was 
made assuming a source spectrum that is  intrinsically flat in $\nu F_\nu$ (i.e., $F_{\nu} \propto \nu^{-1}$, 
a spectral shape very similar to that observed for many distant quasars).  Curves are shown 
for the same four values of comoving distance that are plotted in Fig.~\ref{fig:alpha} 
(here with different line types), and for four discrete values of $\alpha$ (with different colors).  
Beyond the distances where the curves fall off abruptly, any source would be undetectable 
because the light originating 
from the source would be badly out of phase so that formation of an image would be  impossible.
The distant source's photons would simply merge into the noise floor.   
What Fig.~\ref{fig:distmax} shows is that 
while astronomers have only a couple of factors of 10 to work with in distances to AGN, there are
13 orders of magnitude in wavelength from the optical to the TeV $\gamma$-ray range.  This is what
makes the high energy radiation so valuable in constraining $\alpha$. 

\begin{figure}
{\includegraphics[width=1.00\columnwidth]{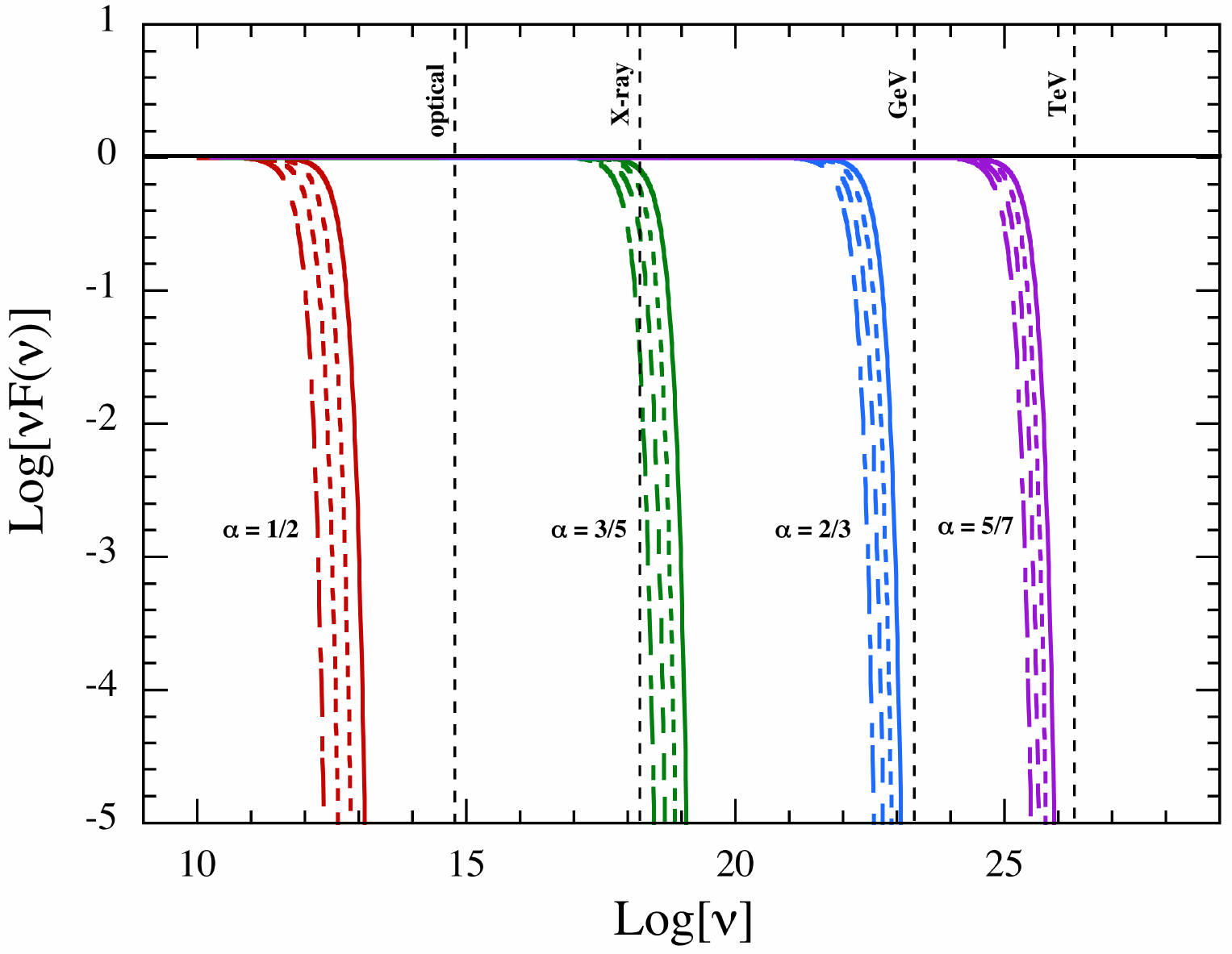}}
\caption{ The relative flux density $\nu F_{\nu}$ for a source, as a function of frequency  $\nu$ for a source at four comoving distances
to the object, respectively 
300 Mpc ($z\approx 0.07$; solid curves), 1 Gpc ($z\approx0.25$; dashed curves), 
3 Gpc ($z\approx 1$; long-dashed curves) and 10 Gpc ($z\approx 12$; dot-dashed curves). As can be seen, for any value of $\alpha$ there is a maximum 
frequency $\nu$ (or, equivalently, a shortest wavelength $\lambda$) beyond which a source would simply be undetectable because the 
phase dispersion for the source's photons would be greater than $\sim 1$ radian, making an image impossible to form.  We have plotted
curves specifically for $\alpha=$1/2 (red), 3/5 (green), 2/3 (blue) and 5/7 (purple).
See \S 3.3
for discussion.}
\label{fig:distmax}
\end{figure}

However, Christiansen et al.~(2011) and Perlman et al.~(2011), as well as earlier workers 
(see previous references) adopted the additional hypothesis that the rms phase fluctuations, 
$\delta \phi$, might also be directly interpretable, to within the same order of magnitude, as 
the diameter of a spacetime foam induced ``seeing disk" for a distant source, $d\psi$.  If that 
were the case, then spacetime foam would have a much more profound effect on the image 
quality by directly blurring the images (see Fig.~1 of Christiansen et al.~2011 and Fig.~1 of 
Perlman et al.~2011), thereby apparently constraining the allowed parameter space to larger 
values of the accumulation factor,  i.e.,  $\alpha  > 0.655$, for optical observations (note: by 
extension, such an interpretation would appear to also allow {\em Chandra} X-ray observations to rule 
out the holographic model (see Fig.~1 of Christiansen et al.~2011).  However, while we can 
construct several scenarios (cf., Christiansen et al.~2006; Christiansen et al.~2011) that 
suggest $\delta \phi \approx \delta \psi$, we do not have a rigorous proof of this hypothesis.   
Because our goal in this paper is to set a definitive limit on $\alpha$ which tests the core 
hypothesis of these models (namely, that spacetime foam directly causes phase fluctuations), 
in this work, we utilize only the more robustly estimated effects of phase fluctuations, which are
{\it independent} of whether the detection device forms an image by reflective or refractive
optics or otherwise (e.g., via the direction of recoiling electrons).  To reiterate, it is the information
carried in the wavefront that determines the best possible image that can be formed, regardless
of the nature or properties of the imager.

In addition to the direct effect of phase fluctuations on the images of distant
astronomical objects, there is also the possibility of  direct deflection
of photons by spacetime foam.  We can construct several dimensional analyses, {\it i.e.},
back-of-the-envelope calculations which might suggest that this  is plausible; 
including possibly photon scattering from 
Planck fluctuations.  However, at this point, these calculations require {\it ad hoc} 
assumptions which go beyond the fundamental theoretical basis of the alpha 
models discussed above. Therefore, in this work, we utilize only the more robustly 
estimated effects of phase fluctuations for setting constraints on the spacetime foam parameter $\alpha$.

\begin{figure*}
\centering
{\includegraphics[width=0.69\textwidth]{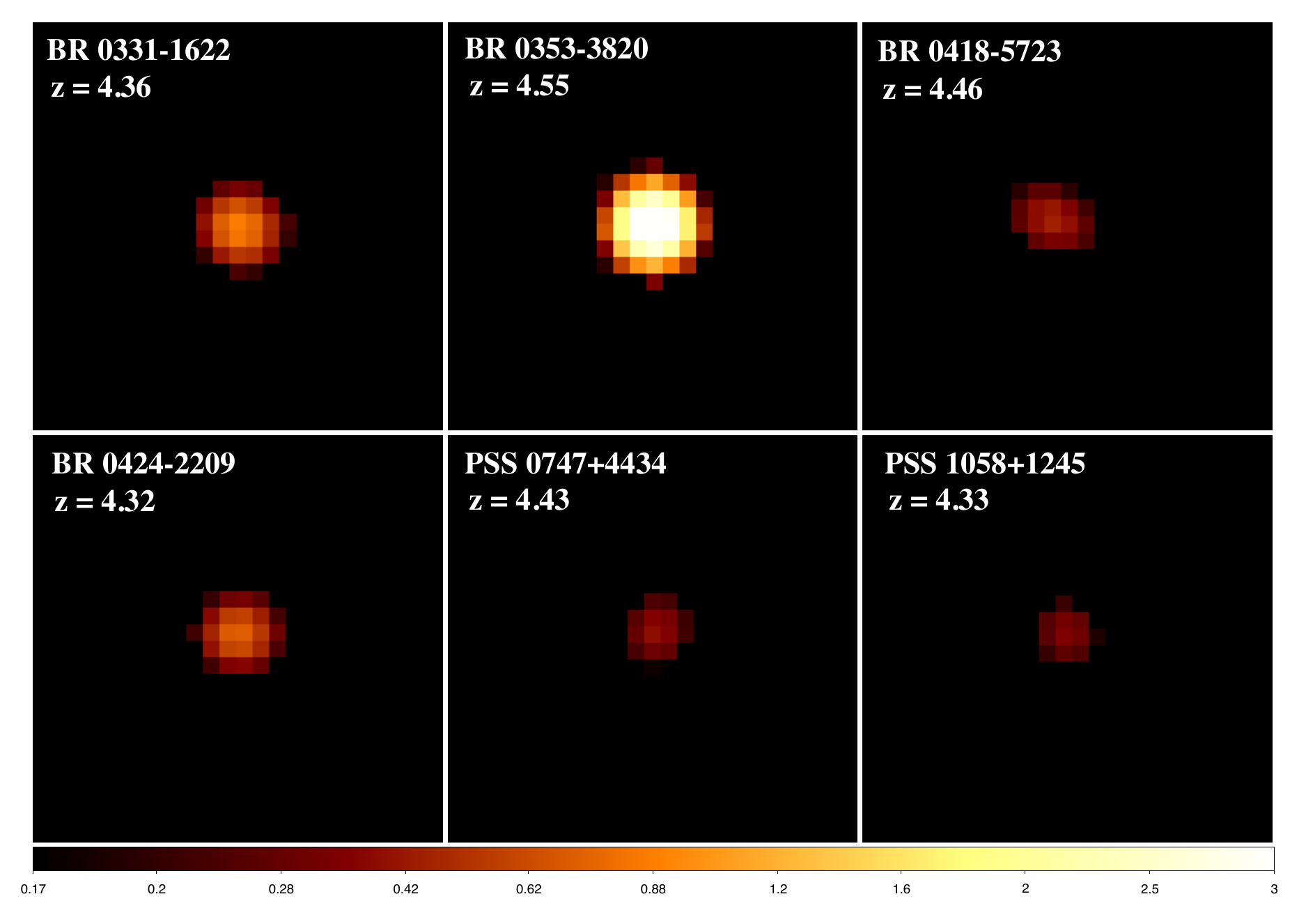}}
\caption{Sample of six {\em Chandra} X-ray images (0.5-8 keV) of quasars with $z > 4$ adapted from Vignali et al.~(2005). The panels are 12$'' \times 12''$ and have been Gaussian smoothed with a 3-pixel ($1.5''$) radius.  The scaling is proportional to the square root of the X-ray flux, and the colorbar indicates counts per smoothed pixel. }
\label{fig:image}
\end{figure*}

\section{Constraints on $\alpha$ From the Existence of Images of Distant High-Energy Sources}
\label{sec:high_energy}

The simulations we have done have profound implications for constraining the 
spacetime foam parameter $\alpha$.   While we have shown that 
optical observations only constrain $\alpha$ to $\alpha>0.53$, rather than the larger values 
found by other authors, we can take advantage of other aspects of the $\alpha$ models
to set tighter constraints.  In particular, equation (\ref{eqn:phirms})
shows that for a given source distance, $\ell$, the rms phase shifts over the wavefront
are proportional to $\lambda^{-1}$.  This opens up the possibility of using X-ray and 
gamma-ray observations to set the tightest constraints yet.   The constraints produced in a 
given band are symbolized in Figs. \ref{fig:alpha} and \ref{fig:distmax} by vertical lines 
that denote optical (5000 \AA~ wavelength or 2.48 eV photon energy), X-ray (5 keV), GeV and TeV 
photons.  The constraints thus produced are lower limits to $\alpha$ produced by the mere observation
of an image (not necessarily a diffraction limited one!) formed of a cosmologically distant source in that
waveband.  Those constraints are summarized in Table 1.

\subsection{Constraints from {\em Chandra} X-Ray Observations}
\label{sec:Xray}

Several dozen high-redshift $(z>2)$ quasars have been observed with {\it Chandra} as the 
specific target of an observation, as well as serendipitously when they happen to be in the same 
field as another object being observed.  

The X-ray images of six very distant (i.e., $z >4$) quasars recorded with the {\em Chandra}
observatory are shown in Figure \ref{fig:image}, taken from the work of Vignali et al.~(2005).  
The sizes of the X-ray images are all consistent with the PSF of the {\em Chandra} X-ray 
optics and demonstrate clearly that the images exist without serious (i.e., orders of magnitude)
degradation in intensity.  This can be inferred, for example, by a comparison of the optical and
X-ray fluxes from these distant quasars, showing that they bear a consistent ratio with observed 
quasars that are much closer to the Earth. Further examples of such work can be found in 
Shemmer et al.~(2006) and Just et al.~(2007). 

From the existence of high quality 
X-ray images of quasars at $z > 4$,
we can constrain $\alpha$ to be $\alpha > 0.58$ (see Figs.~\ref{fig:alpha} and \ref{fig:distmax}, and Table 1).

\subsection{Constraints from Gamma-Ray Observations}
\label{sec:gamma}

For the larger interesting values of $\alpha$ (i.e., 
$\alpha \gtrsim 2/3$, tending to exclude the holographic model), Eqn.~(\ref{eqn:phirms}) indicates 
that for $\ell$ in the range of 100 Mpc to 3 Gpc, the expected phase shifts in the X-ray band 
are only $\lesssim 10^{-4}$ radians.  This is far too small to result in any noticeable effect 
on the X-ray images (unless direct deflection of the X-rays by the spacetime foam is 
possible).  However, for $\gamma$-ray energies ($E_\gamma 
\gtrsim 1$ GeV) the wavelengths are sufficiently short that the phase shifts can exceed $\pi$
radians for $\alpha \simeq 2/3$.  Thus, the mere detection of well-localized $\gamma$-ray
images of distant astronomical objects at wavelengths of of $10^{-13} $ cm or shorter, i.e., photon energies of a GeV 
or higher, will
allow us to place serious constraints on the 
larger values of $\alpha$ (i.e., near to, or greater than, 2/3), thus yielding 
a verdict on the holographic model. 

However, GeV and TeV $\gamma$-rays have the problem that they have wavelengths  
smaller than atomic nuclei, making their detection by geometrical optics 
techniques impossible.  Gamma-ray telescopes rely on the detection of the 
cascades of interactions that happen when $\gamma$-rays impinge on normal matter, whether
the intervening medium be the CsI crystals used in {\it Fermi} (Atwood et al. 2009), or the 
Earth's atmosphere in the TeV (e.g., Aharonian et al. 2008).  In either case the de-coherence of the wave function caused by phase fluctuations, $\delta \phi \sim \pi$,  caused by spactime foam would cause the high energy image to disappear into the noise as discussed in \S 3.2 above.  

\begin{table}
\caption{Constraints on the SpaceFoam Parameter $\alpha$}
\begin{center}
\label{tab:alpha}
\begin{tabular}{cc}
\hline
\hline
\noalign{\smallskip}
waveband              &   lower limit\,$^a$ on $\alpha$   \\
\noalign{\smallskip}
\hline
\noalign{\smallskip}
optical (eV) &   0.53    \\
X-ray (keV)  & 0.58    \\
$\gamma$-rays (GeV)   & 0.67  \\
$\gamma$-rays (TeV)   &  0.72  \\
\noalign{\smallskip}
\hline
\end{tabular}
\end{center}
$^{a}$ See Eqn.~(\ref{eqn:constraint}) and Fig.~\ref{fig:alpha}
\end{table}

The combination of the above suggests that if we can demonstrate the detection of large 
numbers of well localized, cosmologically distant sources in the $\gamma$-ray band (either at GeV or, particularly, 
TeV energies), with reasonable angular resolution, we have a very powerful test of spacetime 
foam models.  For this we start with GeV $\gamma$-rays, where  the dominant extragalactic
sources are distant blazars.  Indeed, the {\it Fermi} gamma-ray space telescope team has firm identifications 
for hundreds of AGN (see Fig.~\ref{fig:Fermi_sky}), over 98\% of which are blazars (Ackermann et al. 2013), 
with redshifts as high as $z=3.2$.  The PSF of Fermi is less than a degree in size for the 1 GeV to 10 GeV 
$\gamma$-rays (Ackermann et al. 2012) and all the blazars are unresolved.  
Two examples  (Mkn 421 at $\ell = 125$ Mpc and 3C 279 at $\ell \simeq 2$ Gpc)of the imaging of active galactic nuclei at a range of energies between
100 MeV and $\gtrsim 100$ GeV with {\em Fermi} are shown in Fig.~\ref{fig:Mkn}. The detection of a large number of GeV 
$\gamma$-ray emitting blazars, sets a constraint of $\alpha>0.67$ on spacetime foam models  (Figs.~\ref{fig:alpha} and \ref{fig:distmax}; Table 1), {\it i.e.}, disfavoring the holographic model, but perhaps not decisively.  

\begin{figure}
\centering
{\includegraphics[width=0.99\columnwidth]{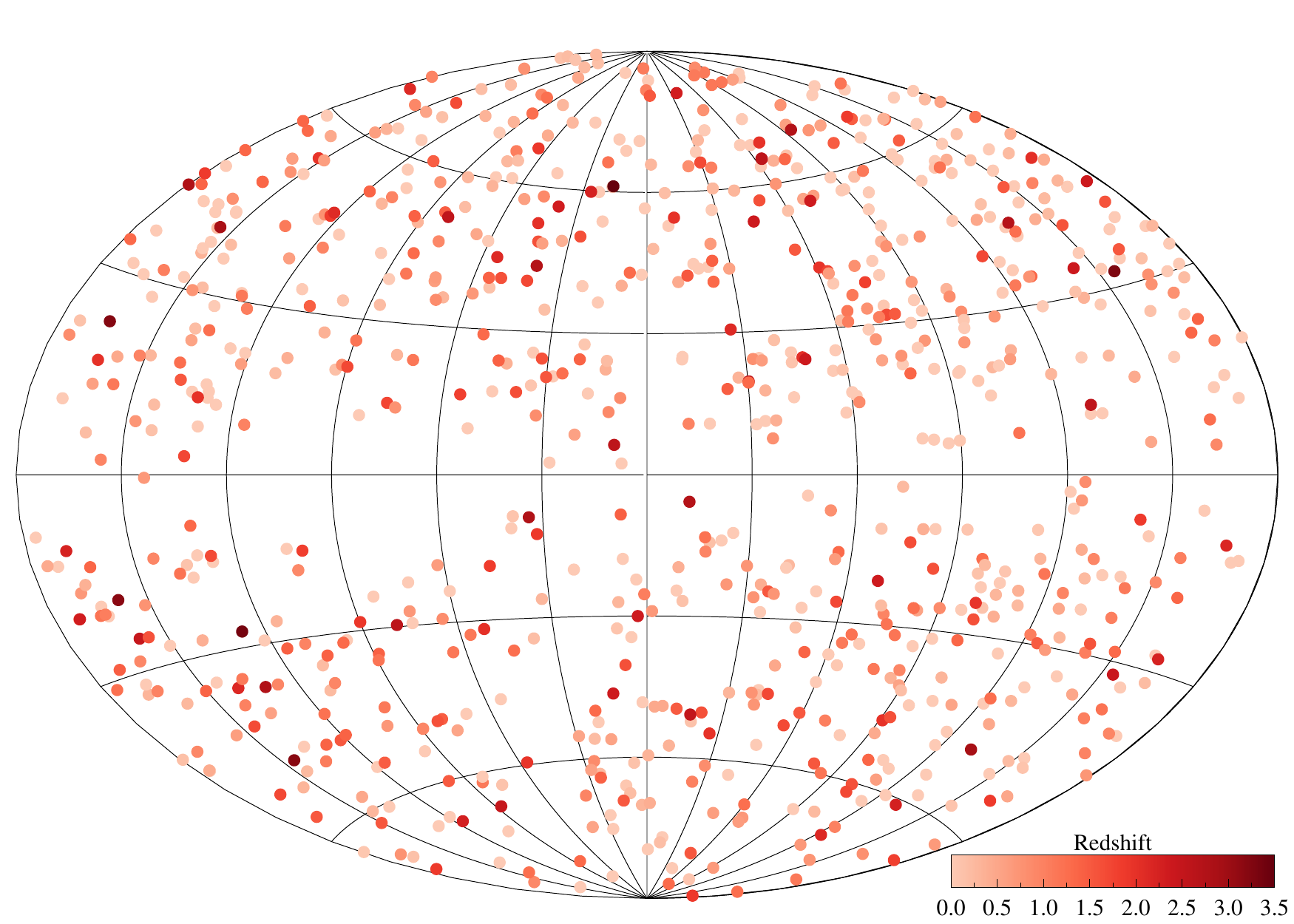}}
\caption{The locations of {\em Fermi} extragalactic $\gamma$-ray sources in the GeV band.  From Ackermann et al.~(2011).  The intensity of the circles is an indication of the redshift.}
\label{fig:Fermi_sky}
\end{figure}

\begin{figure}
\centering
{\includegraphics[width=1.00\columnwidth]{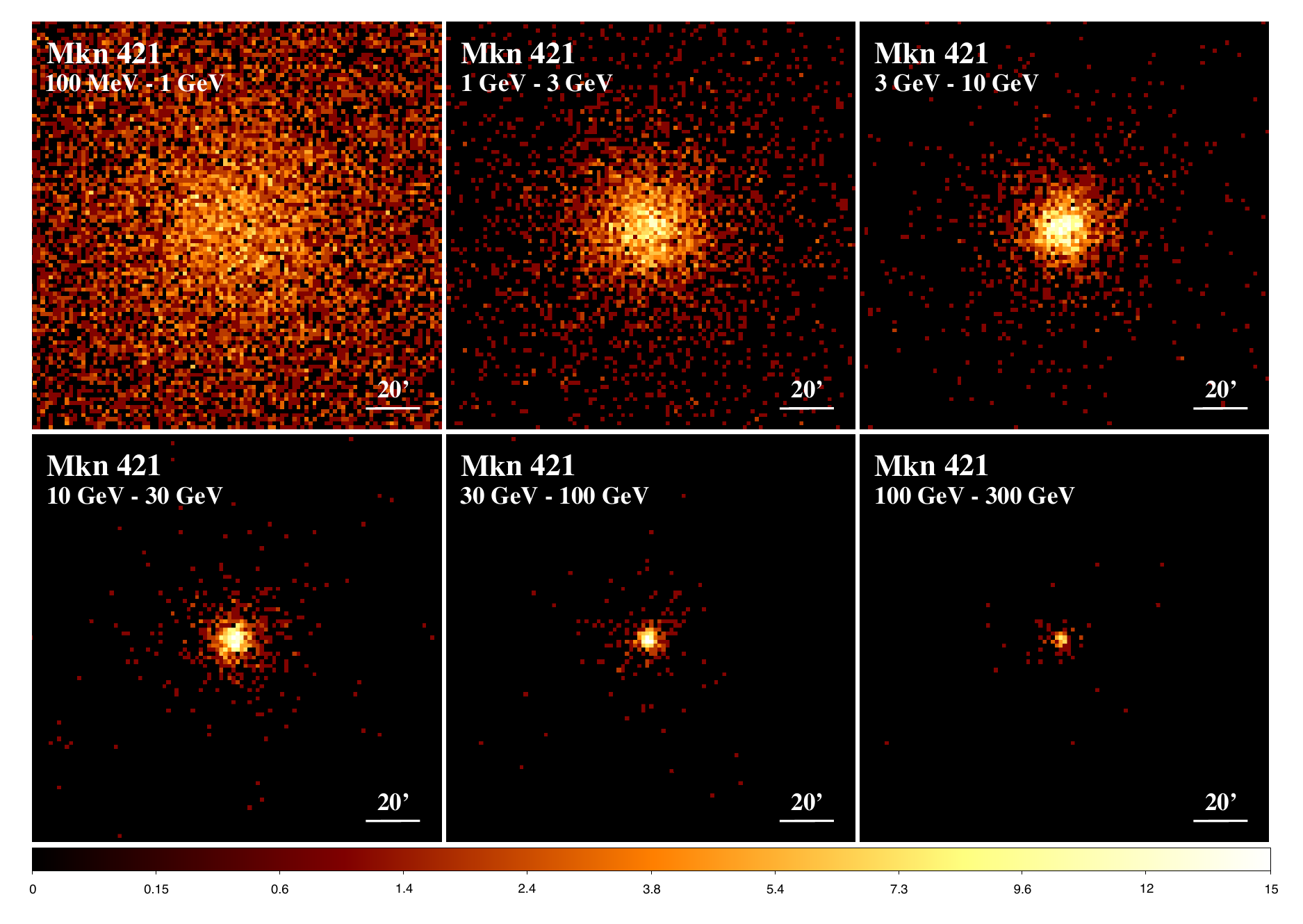}}
{\includegraphics[width=1.00\columnwidth]{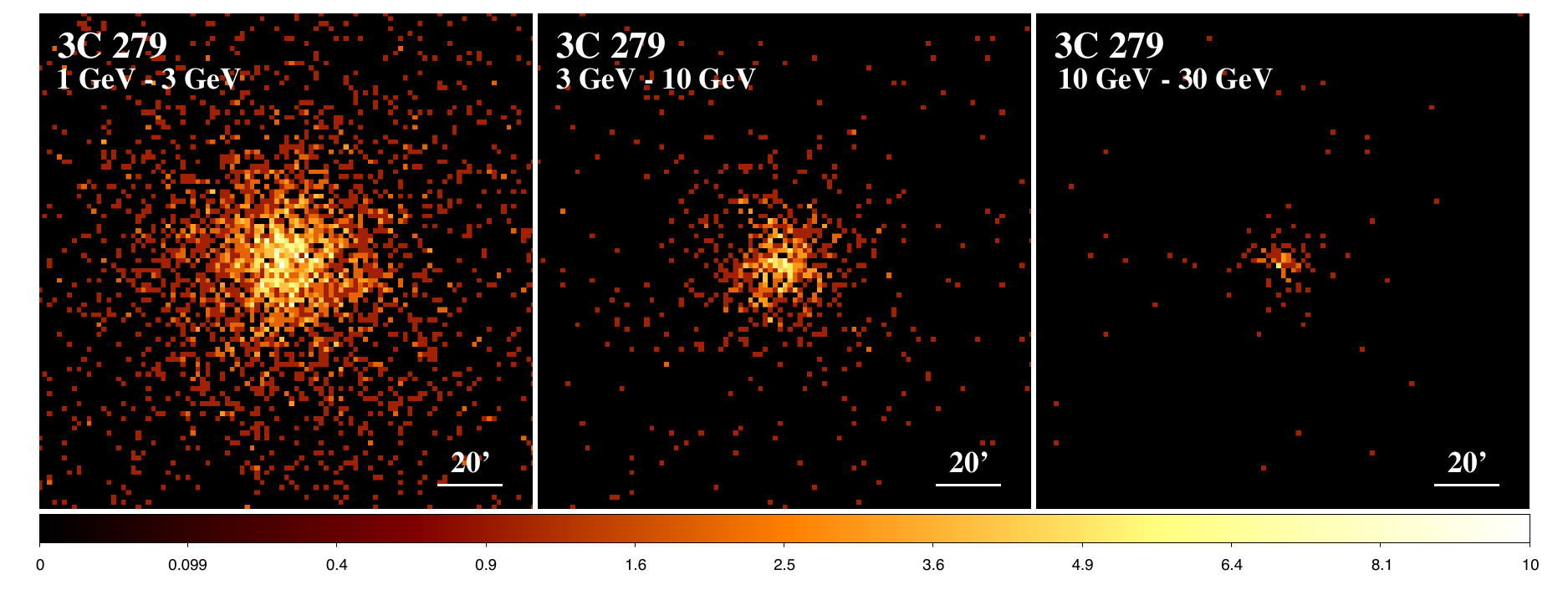}}
\caption{Top panels: {\em Fermi} images of Mkn 421 ($z=0.034, \ell = 125$ Mpc)
in six different energy bands ranging from $\sim$100 MeV to 100 GeV.  In all cases, above $\sim$1 GeV, the image is well formed and localized to better than $\sim$20$'$, especially at the higher energies. Bottom panels: {\em Fermi} images of 3C 279  ($z=0.536, \ell$ = 2 Gpc) in three different energy bands ranging from $\sim$1 GeV to 30 GeV.}
\label{fig:Mkn}
\end{figure}

At higher energies (i.e., TeV), there are several telescopes and telescope arrays.  For example, 
VERITAS is an array of four 12m-diameter imaging atmospheric Cherenkov telescopes located
at the Fred Lawrence Whipple Observatory in southern Arizona.  VERITAS is designed to 
measure photons in the energy range 100 GeV to 30 TeV with a typical energy resolution of 
15-20\%.  VERITAS features an angular resolution of about $0.1^\circ$ in a $3.5^\circ$ field of 
view (Holder et al.~2006, Aharonian et al.~2008, Kieda et al.~2013).  
The performance characteristics of VERITAS are reasonably similar to those of the other 
major TeV arrays (e.g., HESS, Giebels et al.~2013; MAGIC, Aleksic et al.~2014a,b) in terms of angular 
resolution.  Together, the TeV telescopes have detected 55 extragalactic sources, all but three of 
which are distant blazars{\footnote{see http://tevcat.uchicago.edu, and references therein}}.  The 
highest redshift source to have been detected in TeV $\gamma$-rays is S3 0218+35, a 
gravitationally lensed blazar at $z=0.944$ (Mirzoyan et al.~2014).  All of the extragalactic sources 
known are unresolved with the TeV telescopes.

The detection of  distant, TeV $\gamma$-ray emitting 
blazars, sets a constraint of $\alpha>0.72$  (Figs.~\ref{fig:alpha} and \ref{fig:distmax}; Table 1)
on spacetime foam models, i.e., strongly 
disfavoring, if not altogether ruling out the holographic model.

\section{Summary and Conclusions}
\label{sec:summary}

In this work we have discussed how spacetime foam can introduce small scale 
fluctuations in the wavefronts of distant astronomical objects. We have shown that 
when the path-length fluctuations in the wavefront become comparable to the wavelength
of the radiation, the images will basically disappear.  Thus, the very {\em existence} of distant
astronomical images can be used to put significant constraints on models of spacetime foam.

The existence of clear, sharp (i.e., arc-second) {\em Chandra} X-ray images of distant
AGN and quasars, at intensities that are not very far from what is expected based on 
similar objects at closer distances, tells us that the parameter $\alpha$ must exceed
0.58.  This rules out the so-called ``random walk'' model ($\alpha = 1/2$; see 
\S\ref{sec:phase_fluc}).

Perhaps the strongest constraints of all now come from the detection of large numbers 
of cosmologically distant sources -- mostly blazars -- in the $\gamma$-rays.  
These detections limit $\alpha$ to values higher than 0.67 and 0.72, at GeV and TeV energies, 
respectively (see Figs.~\ref{fig:alpha} and \ref{fig:distmax} as well as Table 1).  
This strongly disfavors, if not completely rules out, the holographic model.

We should recall that the 
spacetime foam model parametrized by $\alpha = 2/3$, as formulated
(Ng \& van Dam, 1994, 1995), is called the `holographic 
model' only because it is {\it consistent} (Ng 2003) with the holographic
principle; the demise of the model may {\it not} necessarily imply the demise
of the principle since it is conceivable that the correct spacetime 
foam model associated with the holographic principle can take on a 
different and more subtle form than that which can be
given by $\delta \ell \approx \ell^{1/3} \ell_P^{2/3}$.  
It is important to be clear: what we are ruling out (subject to the
caveat mentioned above) are the models with
$\alpha < 0.72$ for the spacetime foam models that can be categorized
according to $\delta \ell \approx \ell^{1 - \alpha} \ell_P^{\alpha}$. 

On the other hand, it is legitimate to ask what, if any, is (are) the implication(s) that 
the  $\alpha = 2/3$ spacetime foam model is indeed ruled out.  We recall that, aside 
from simple quantum mechanics, essentially the only ingredient that has been used 
(Ng \& van Dam 1994; 1995; see also \S \ref{sec:holog}) in the derivation of the result 
$\delta \ell \approx \ell^{1/3} \ell_P^{2/3}$ is the requirement that the mass ($M$) and size 
($\ell$) of the system under consideration satisfy $M < \ell c^2/2G$ because we need 
information about the system to be observable to outside observers.  Now one way 
that this requirement can be waived is that gravitational collapse produces apparent 
horizons but no event horizons behind which information is lost, which has 
recently been proposed by Hawking (Hawking 2014); see also Mersini-Houghton
(2014).  It is tempting to interpret our result that the spacetime foam model for which 
$\alpha = 2/3$ is ruled out as the first albeit indirect observational affirmation of the 
idea that gravitational collapse indeed does not necessarily produce an event horizon.

\begin{acknowledgments}

YJN was supported in part by the Bahnson Fund of the University of
North Carolina at Chapel Hill.  We thank A.  Glindemann for very useful 
discussions regarding the effects of phase fluctuations on the images.  We also thank an 
anonymous referee for useful comments that significantly improved this paper.
 
\end{acknowledgments}

\end{document}